\documentclass{article}

\usepackage[english]{babel}
\usepackage{bm}
\usepackage[fleqn]{amsmath}
\usepackage{amsfonts}
\usepackage{amsgen,amsthm}
\usepackage{amssymb}
\usepackage{amsbsy}
\usepackage{bbm}
\usepackage{graphicx}
\usepackage{booktabs}
\usepackage{dcolumn} 
\usepackage[a4paper,margin=3cm]{geometry}
\usepackage{appendix}
\usepackage{lastpage}
\usepackage{enumerate}
\usepackage{textcomp}
\usepackage{calc}
\usepackage{ifthen}
\usepackage[square,numbers,compress]{natbib}
\usepackage{accents}
\newlength{\dhatheight}

\providecommand\der{{d}}
 
\newcommand{\aderv}[2]{\frac{\partial {#1}}{\partial {#2}}}
\newcommand{\adervo}[2]{\frac{\der {#1}}{\der {#2}}}

\newcommand{\adervs}[2]{\frac{\partial^2 {#1}}{\partial {#2}^2}}
\newcommand{\adervso}[2]{\frac{\der^2 {#1}}{\der {#2}^2}}

\newcommand{\equa}[1]{Eq.~(\ref{#1})}

\newcommand{\equas}[1]{Eqs.~(\ref{#1})}
\newcommand{\equass}[2]{Eqs.~(\ref{#1})--(\ref{#2})}
\newcommand{\equasa}[2]{Eqs.~(\ref{#1}){ }and{ }(\ref{#2})}
\renewcommand{\Re}{\mathfrak{Re}}

\newcommand{\eqn}[2]{\begin{gather}
#1
\label{#2}
\end{gather}
}

\newcommand{\spl}[2]{\begin{gather}
\displaybreak[2]
\begin{split}
#1
\end{split}
\label{#2}
\end{gather}
}

\newcommand{\gat}[2]{\begin{subequations}\label{#2}\begin{gather}
#1
\end{gather}\end{subequations}
}

\numberwithin{equation}{section}

\title{\bf Transition to Absolute Instability\\
for ({\em not so}) Dummies}
\author{\textbf{\em Antonio Barletta}  \\
Department of Industrial Engineering \\[-6pt] 
Alma Mater Studiorum Universit\`a di Bologna\\[-6pt]   
Viale Risorgimento 2, 40136 Bologna (Italy)\\[-6pt] 
\texttt{antonio.barletta@unibo.it}\\[6pt]
\textbf{\em Leonardo Santos de Brito Alves}  \\
Universidade Federal Fluminense\\[-6pt]
Centro Tecnol\'ogico -- Escola de Engenharia\\[-6pt]
Rua Passo da P\'atria, 156, bloco E, sala 216\\[-6pt]
S\~ao Domingos, 24210-240 -- Niteroi, RJ (Brazil)\\[-6pt] 
\texttt{leonardo.alves@mec.uff.br}
	}

\begin{document}

\maketitle

\vspace{0.5cm}

\section*{Foreword}
These notes are intended as an elementary introduction to the concept of absolute instability. The transition from convective instability to absolute instability is an important issue when the stability of stationary flow solutions is investigated. The arguments here described were first developed in the framework of plasma physics and later applied to the hydrodynamics of mixing layers and shear flows. Far from being a comprehensive analysis of this complicated subject, the aim of these notes is just to sketch a rudimentary and quite elementary ground for physicists or engineers which have a familiarity with the basic features of linear stability in fluid dynamics, but are new to the concept of absolute instability.

\tableofcontents

\section{Introduction}
The idea behind the distinct concepts of convective instability and absolute instability dates back to some decades ago. The origin is in the sixties of the past century, within the area of plasma physics. Early studies on this topics are, among the others, the papers by Dysthe \cite{dysthe1966} and Gaster \cite{gaster1968}. A description of the transition to absolute instability in fluid dynamics is also available in Chapter III, Section 28, of Landau \& Lifschitz \cite{landau1987}. The modern approach to this subject was developed starting from the eighties \cite{huerre1985, monkewitz1990, brevdo1991, chomaz1992, carriere1999, delache2007}. 

This elementary description aims to focus a few simple physical aspects of the dualism convective instability versus absolute instability, and to fix some basic ideas on the method to detect the onset of both types of instability. The forthcoming analysis has no intent of being comprehensive or rigorous. For both these tasks, we warmly suggest the reader to study the exhaustive reviews by Huerre \& Monkewitz \cite{huerre1990}, and by Huerre \cite{huerre2000}, as well as the textbooks by Criminale et al. \cite{criminale2003} (see Section 4.4),  by Drazin \& Reid \cite{drazin2004} (see Section 24), and by Charru \cite{charru2011} (see Chapter 3). Fresh new insights into this topic have been also proposed in the recent paper by \'O N\'araigh \& Spelt \cite{onaraigh2013}.

\section{Wave packets and their time evolution}
A perturbation of a stationary flow can be reasonably described as localised in space, at least at the beginning of its time evolution. This nature of the actual perturbations acting on a flow system is in contrast with the usual analysis of linear stability based on normal modes. In fact, normal modes are plane waves and, hence, they are definitely non-local as they are spread over all accessible space.
Configurations of waves may exist where localisation emerges in the form of wave packets. The basic mathematical theory is the Fourier analysis. 

\subsection{Stationary waves, $x$--space and $k$--space}
One may imagine a function $f(x)$ expressed as
\eqn{
f(x) = \int_{-\infty}^{\infty} dk\ g(k) e^{i k x} .
}{8}
\equa{8} defines $f(x)$ as a linear superposition of infinite standing plane waves with wavelength $\lambda = 2\pi/k$. Two neighbouring maxima of the real and imaginary parts of $e^{i k x}$ are separated by a distance $2\pi/k$. Each wave is ``weighted'' by the coefficient function $g(k)$. \\
One may consider a Gaussian weight function
\eqn{
g(k) = e^{-\alpha \left( k - k_0 \right)^2} ,
}{9}
where $\alpha > 0$. One can substitute \equa{9} into \equa{8} and evaluate the integral on the right hand side of \equa{8},
\eqn{
f(x) = \int_{-\infty}^{\infty} dk\ e^{-\alpha \left( k - k_0 \right)^2} e^{i k x} = e^{i k_0 x}  \int_{-\infty}^{\infty} dk'\ e^{-\alpha k'^2} e^{i k' x} = e^{i k_0 x} \sqrt{\frac{\pi}{\alpha}} \ e^{-x^2/(4 \alpha)} .
}{10}
On considering the square modulus of $g(k)$ and the square modulus of $f(x)$, 
\eqn{
|g(k)|^2 = e^{- 2 \alpha \left( k - k_0 \right)^2} , \qquad |f(x)|^2 =  \frac{\pi}{\alpha} \ e^{-x^2/(2 \alpha)} ,
}{11}
one realizes that we have a Gaussian signal both in $k$--space and in $x$--space. We may easily check that, when $k = k_0 \pm \Delta k/2$, where $\Delta k = 2/\sqrt{2 \alpha}$, the Gaussian signal in $k$--space drops to $1/e$ times its peak value. When $x = \pm \Delta x/2$, where $\Delta x = 2\sqrt{2\alpha}$, the Gaussian signal in $x$--space drops to $1/e$ times its peak value. If $\alpha$ becomes smaller and smaller, the signal in $k$--space increases its width $\Delta k$, while the signal in $x$--space decreases its width $\Delta x$. One may easily check that  
\eqn{
\Delta k \ \Delta x = 4 .
}{12}
The precise numerical value of the product is not important. What is important is that the product $\Delta k \ \Delta x$ is finite and independent of $\alpha$. A highly localised wave packet in $k$--space, i.e. one with a small $\Delta k$, means a poorly localised wave packet in $x$--space, i.e. one with a large $\Delta x$, and vice versa. 
\begin{figure}[h!]
\begin{center}
\includegraphics[width=0.8\textwidth]{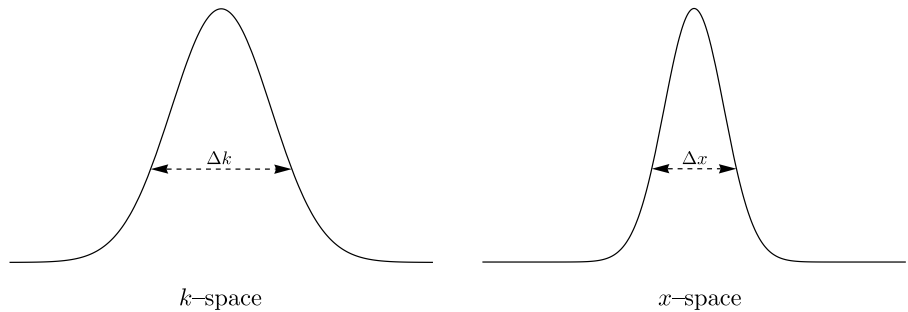}
\end{center}
\caption{Localization of wave packets in $k$--space and in $x$--space}
\label{fig1}
\end{figure}
It is not possible to reduce the width of the Gaussian signal both in $k$--space and in $x$--space.

\subsection{Wave packets}
Let us now consider a perturbation wave packet made up with the superposition of travelling plane waves. Then, \equa{8} is to be replaced with
\eqn{
F(x,t) = \int_{-\infty}^{\infty} dk\ g(k) e^{i \left( k x - \omega t \right)} ,
}{14}
where $\omega$ is the angular frequency, while $k$ is the wave number. The simplest case is that with $\omega = c k$, where $c$ is a constant. In this case, \equa{14} describes a superposition of plane waves having a constant phase velocity $c$. Then, a comparison between \equa{8} and \equa{14}, allows one to write
\eqn{
F(x,t) = f(x - c t) .
}{15}
The effect of the standing waves being replaced by travelling waves is just a rigid translational motion of the wave packet with a velocity $c$. No distortion of the wave packet is caused by the time evolution.

\subsection{Group velocity and spreading}\label{gro}
The linear stability analysis of flow systems leads generally to situations where $\omega$ is not simply proportional to $k$. We are therefore interested in assuming a general relationship $\omega = \omega(k)$. We may assume a wave packet strongly localised in $k$--space, with a marked peak at $k=k_0$, expressed by \equa{9}, namely a {\em quasi-monochromatic} wave packet.
 
The strong localisation in $k$--space suggests that one may express $\omega(k)$ as a Taylor expansion around $k=k_0$ truncated to lowest orders,
\eqn{
\omega(k) \approx \omega(k_0) + \left. \adervo{\omega}{k} \right|_{k=k_0}\!\! \left( k - k_0 \right) + \frac{1}{2} \left. \adervso{\omega}{k} \right|_{k=k_0}\!\! \left( k - k_0 \right)^2 +\ \ldots \ .
}{16}
We use the notations
\eqn{
v_g = \left. \adervo{\omega}{k} \right|_{k=k_0} , \qquad \beta = \frac{1}{2} \left. \adervso{\omega}{k} \right|_{k=k_0} ,
}{17}
where $v_g$ is called the \emph{group velocity}.\\
We now substitute \equasa{9}{16} in \equa{14} and we obtain
\spl{
F(x,t) &= \int_{-\infty}^{\infty} dk\ e^{-\alpha \left( k - k_0 \right)^2} e^{i \left\{ k x - [\omega(k_0) + (k - k_0) v_g + (k - k_0)^2 \beta ] t \right\}} \\
&= \int_{-\infty}^{\infty} dk'\ e^{-\alpha k'^2} e^{i \left\{ (k_0 + k') x - [\omega(k_0) + k' v_g + k'^2 \beta ] t \right\}} \\
&= e^{i \left[ k_0  x - \omega(k_0) t \right]} 
\int_{-\infty}^{\infty} dk'\ e^{- \left( \alpha + i \beta t \right) k'^2} e^{i k' \left( x - v_g t \right)} 
\\
&= e^{i \left[ k_0  x - \omega(k_0) t \right]} 
\int_{-\infty}^{\infty} dk'\ e^{- \alpha' k'^2} e^{i k' x'} ,
}{18}
where $\alpha' = \alpha + i \beta t$ and $x' = x - v_g t$. We note that the integral appearing in \equa{18} is just the same as that evaluated in \equa{10}, with $\alpha$ replaced by $\alpha'$ and $x$ replaced by $x'$. Thus, we may write
\spl{
F(x,t) &= e^{i \left[ k_0  x - \omega(k_0) t \right]} 
\int_{-\infty}^{\infty} dk'\ e^{- \alpha' k'^2} e^{i k' x'} \\
&= e^{i \left[ k_0  x - \omega(k_0) t \right]} \sqrt{\frac{\pi}{\alpha + i \beta t}} \ e^{- \left( x - v_g t \right)^2/\left[4 \left( \alpha + i \beta t \right) \right]} .
}{19}
Again, we consider the square moduli of $g(k)$ and of $F(x,t)$ as in \equa{11}, 
\eqn{
|g(k)|^2 = e^{- 2 \alpha \left( k - k_0 \right)^2} , \qquad |F(x,t)|^2 =  \frac{\pi}{\sqrt{ \alpha^2 + \beta^2 t^2 }} \ e^{- \alpha \left( x - v_g t \right)^2/\left[2 \left( \alpha^2 + \beta^2 t^2 \right) \right]} .
}{20}
One recognizes Gaussian signals both in $k$--space and in $x$--space. The peak of the Gaussian signal in $x$ is located at $x = v_g t$, and thus it travels in the $x$--direction with the constant group velocity, $v_g$.

The width of the Gaussian signal in $k$--space is obviously $\Delta k = 2/\sqrt{2 \alpha}$, while the width of the Gaussian signal in $x$--space is now a function of time, 
\eqn{
\Delta x = 2 \sqrt{2 \alpha}\ \sqrt{1 + \frac{\beta^2 t^2}{\alpha^2}} .
}{21}
The width of the Gaussian signal in $x$--space increases in time. This means that the time evolution of the wave packet implies a spreading in $x$--space with a decreasing value at the peak position, $x = v_g t$. The latter feature can be easily inferred from \equa{20}.
\begin{figure}[h!]
\begin{center}
\includegraphics[width=0.8\textwidth]{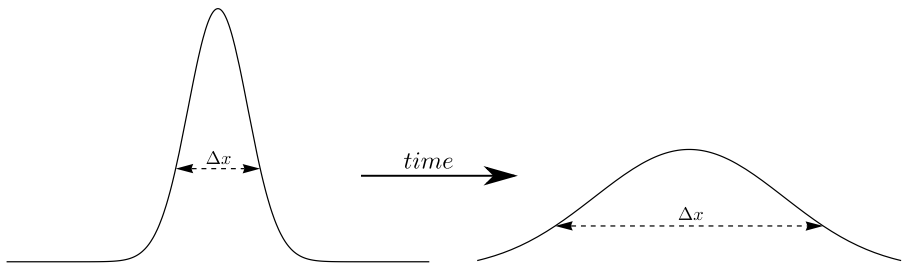}
\end{center}
\caption{Spreading of the wave packet in $x$--space}
\label{fig2}
\end{figure}\\

\section{Convective to absolute instability}
When wave packets are used to model perturbations in a linear stability analysis, the time evolution is modulated by an additional time-dependent factor in the amplitude. This effect might lead to instability if the time evolution is such that the spreading of the wave packet is accompanied by a growth in amplitude. Stability, on the contrary arises when the time evolution is such that the spreading of the wave packet is accompanied by a decrease in amplitude.

\subsection{Temporal and spatial analyses of stability}
This modification in the evolution of wave packets with respect to what has been described in the preceding section is usually accounted for by reconsidering the normal modes used in \equa{14}. Now, the modes packed to form the perturbation $F(x,t)$ are given by
\eqn{
\psi(x,t) = \hat{\psi}\ e^{i \left( k x - \omega t \right)} ,
}{22}
where, now, $k = k_r + i k_i \in \mathbb{C}$, and $\omega = \omega_r + i \omega_i \in \mathbb{C}$. Two main cases are usually studied.\\[6pt]
(1)\qquad{}{\em Temporal stability analysis} $(k_i = 0, \mbox{or}\ k \in \mathbb{R})$. It is the classical basis for a linear stability analysis. This choice is suitable for the determination of the marginal stability condition and of the critical value of the order parameter for the transition to convective instability.\\[6pt]
(2)\qquad{}{\em Spatial stability analysis} $(\omega_i = 0, \mbox{or}\ \omega \in \mathbb{R})$. It is the choice to be preferred for the determination of the transition from convective to absolute instability and of the threshold value of the order parameter for the transition to absolute instability. The role of $k_i$ is to establish if the normal mode grows when $x \to +\infty$, i.e. when $k_i < 0$, or if it grows when $x \to -\infty$, i.e. when $k_i > 0$. The case $k_i = 0$ describes a plane wave with constant and uniform amplitude.

\subsection{Convective versus absolute nature of instability}
\begin{figure}[h!]
\begin{center}
\includegraphics[width=0.3\textwidth]{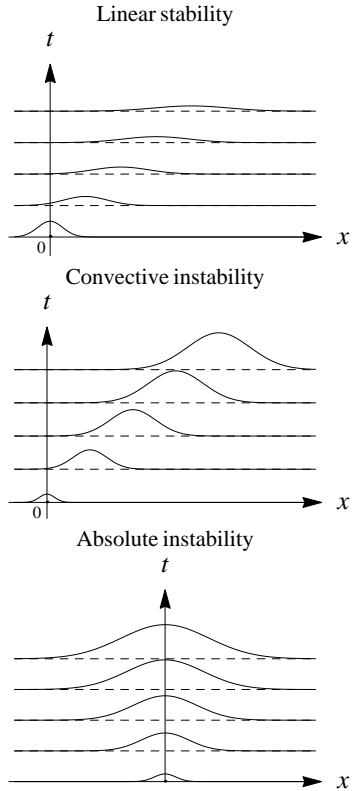}
\end{center}
\caption{Behaviour of wave packet perturbations under stability, convective instability, or absolute instability conditions}
\label{fig3}
\end{figure}
Having anticipated the terms ``convective instability'' and ``absolute instability'', it is time to define them properly. 

The evolution of a perturbation wave packet can be described according to different inertial reference frames. Each reference frame corresponds to a ray,
\[
\frac{x}{t} = constant ,
\]
where the constant can be any real value. Strictly speaking, all rays are equivalent but, for practical reasons, the ray $x/t = 0$ is special as it models a measurement device in a laboratory, i.e. the ray $x/t = 0$ defines the laboratory reference frame.
\\
Let $F(x,t)$ be an arbitrary wave packet describing a localised perturbation of the basic flow. We may have three conditions:\\[6pt]
\textbf{\em Linear stability}
\eqn{
\lim_{t\to \infty} |F(x,t)|^2 = 0, \quad \mbox{along every ray}\quad \frac{x}{t} = constant .
}{23}
~\\[-6pt]
\textbf{\em Convective instability}\\
There exists at least one ray
\eqn{
\frac{x}{t} = constant , \quad \mbox{such that} \quad \lim_{t\to \infty} |F(x,t)|^2 = \infty  .
}{24}
~\\[-6pt]
\textbf{\em Absolute instability}
\eqn{
\lim_{t\to \infty} |F(x,t)|^2 = \infty, \quad \mbox{along the ray}\quad \frac{x}{t} = 0 .
}{25}
~\\[-6pt]
An illustration of what \equass{23}{25} define is given in Fig.~\ref{fig3}. We can say that a convectively unstable basic flow amplifies a given perturbation and convects it away, so that the time growth of the amplitude is not actually observed in the laboratory reference frame. On the other hand, an absolutely unstable flow is amplified in place, and its unbounded amplitude growth is measurable in the laboratory reference frame, i.e. along the ray $x/t=0$.

\subsection{Rays and group velocity}
We learned in Subsection~\ref{gro} that the peak of a localised wave packet travels along the $x$--axis with a speed given by the group velocity,
\eqn{
v_g = \left. \adervo{\omega}{k} \right|_{k=k_0} .
}{26}
Then, an important ray for judging the stable or unstable behaviour of the wave packet is
\eqn{
\frac{x}{t} = v_g  .
}{27}
If we now refer to the general normal modes given by \equa{22} under a temporal stability analysis, where $k \in \mathbb{R}$ and $\omega = \omega_r + i \omega_i \in \mathbb{C}$, we are ready to express a sufficient condition for absolute instability:
\eqn{
v_g = \left. \adervo{\omega_r}{k} \right|_{k=k_0} = 0, \qquad \omega_i(k_0) > 0,
}{27}
where $\omega_i$ describes either an exponential time growth or damping depending on its sign. If $k_0$ satisfies \equa{27}, then $\omega_i(k_0)$ is termed \emph{absolute growth rate}.

\subsection{Absolute instability and the spatial modes}
In a linear stability analysis, the perturbation $F(x,t)$ must be a solution of a linear differential equation,
\eqn{
\mathcal{D}\Big(- i\, \aderv{}{x},\ i\, \aderv{}{t} \Big) F(x,t) = 0 .
}{28}
On claiming the validity of \equa{28} for the normal modes \equa{22}, one is lead to the \emph{dispersion relation},
\eqn{
\mathcal{D}(k, \omega ) = 0 .
}{29}
The dispersion relation may be studied for the spatial modes, with $k = k_r + i k_i \in \mathbb{C}$ and $\omega \in \mathbb{R}$. In fact, \equa{29} is a complex-valued equation that yields two real equations, one for the vanishing real part of the left hand side, and one for the vanishing imaginary part of the left hand side. The former equation generally does not contain $\omega$ when the governing equation of fluid flow is of first-order in time. This is the case for Newtonian and non-Newtonian purely viscous or viscoplastic fluids, while this condition does not apply in the case of viscoelasticity. Hence, one may write
\eqn{
\Re\{\mathcal{D}(k, \omega )\} = G(k_r, k_i ) = 0 .
}{30}
Further governing parameters may appear in \equa{30} not explicitly mentioned, say the Reynolds number or the Rayleigh number. Thus, on varying these parameters, one may figure out \equa{30} as the equation of a bundle of curves in the $(k_r, k_i)$--plane. When the governing parameters define a condition of stability, the curve given by \equa{30} generally features two disconnected branches: one entirely laying in the half-plane $k_i > 0$, and one entirely laying in the half plane $k_i < 0$. The former branch defines spatial modes unboundedly growing upstream, i.e. in the limit $x \to -\, \infty$, the latter branch is made up of spatial modes exponentially growing downstream. When one of the two branches becomes tangent to the line $k_i=0$, this marks the onset of convective instability and $k_r$ becomes equal to the critical wave number, $k_c$. A possibility is that the onset of convective instability takes place with the two disconnected branches merging at a point on the line $k_i = 0$, called the \emph{pinching point}. If this the case, the onset of convective instability is also the onset of absolute instability, i.e. the two conditions coincide.

In general, convective instability and absolute instability are distinct conditions. As a consequence, the pinching point for the two branches of spatial modes in the $(k_r, k_i)$--plane occurs either in the half plane $k_i > 0$ or in the plane $k_i < 0$. In any case, it is the merging of the two branches that defines the onset of absolute instability. In fact, the mixing up of modes growing both upstream and downstream  suggests that the disturbance is undamped and is not swept away by the basic flow. The pinching point in the $(k_r, k_i)$--plane defines a condition of vanishing group velocity. The exact value of the governing parameters that leads to the transition to absolute instability, and to the pinching point for the two branches in the $(k_r, k_i)$--plane, can be in fact determined by solving
\spl{
&\mathcal{D}(k, \omega ) = 0 ,\\
&\aderv{}{k} \mathcal{D}(k, \omega ) = 0 , \quad \mbox{with} \quad \adervo{\omega}{k} = 0 .
}{31}

\section{A neat example: Burgers' equation}\label{burg}
A very simple example to test the description provided in the preceding section is given by the stability analysis of the one-dimensional Burgers' equation,
\eqn{
\aderv{U}{t} + U\ \aderv{U}{x} = \adervs{U}{x} + F(U) ,
}{32}
where we assume a linear $U$--dependent force term,
\eqn{
F(U) = R \left( U - U_0 \right) , \qquad R \in \mathbb{R}, \quad U_0 \in \mathbb{R} . 
}{33}
Evidently $U = U_0$ is a basic solution of \equa{32}, whatever is the value of the real constant $U_0$.

\subsection{The dispersion relation}
A linear stability analysis of this constant solution, $U = U_0$, is carried out by perturbing it, namely by substituting
\eqn{
U = U_0 + \varepsilon\, u , \qquad \varepsilon > 0 
}{34}
into \equa{32} and by taking into account \equa{33},
\eqn{
\aderv{u}{t} + U_0\ \aderv{u}{x} + \varepsilon\, u\ \aderv{u}{x} = \adervs{u}{x} + R\, u .
}{35}
Here, $\varepsilon$ is a perturbation parameter such that $\varepsilon \ll 1$. Then, \equa{35} is linearised to yield
\eqn{
\aderv{u}{t} + U_0\ \aderv{u}{x} = \adervs{u}{x} + R\, u .
}{36}
\begin{figure}[t]
\begin{center}
\includegraphics[width=0.85\textwidth]{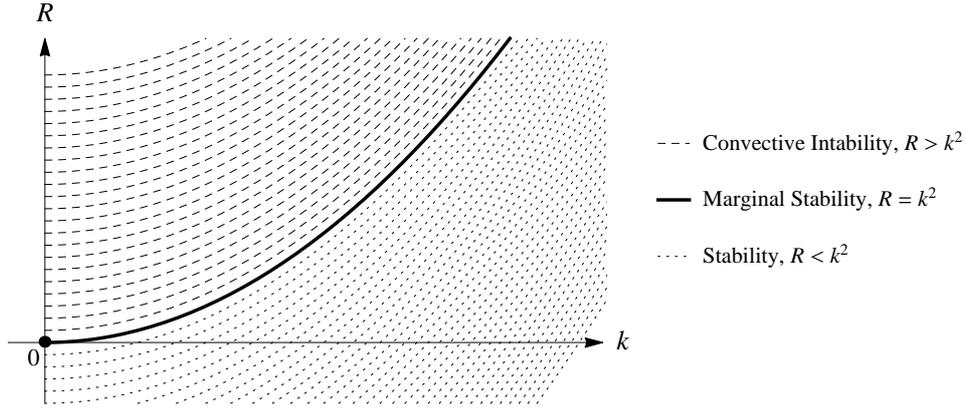}
\end{center}
\caption{Temporal modes: transition from stability to convective instability}
\label{fig4}
\end{figure}

\noindent We now substitute the normal modes \equa{22} into \equa{36}, so that one is lead to the dispersion relation,
\eqn{
\mathcal{D}(k, \omega ) = -\, i \left( \omega - U_0 k \right) + k^2 - R = 0 .
}{37}

\subsection{Temporal modes}
If we perform a temporal stability analysis, then $k \in \mathbb{R}$ and $\omega = \omega_r + i \omega_i \in \mathbb{C}$. We separate the real from the imaginary part in \equa{37}, so that we get
\gat{
R = k^2 + \omega_i , \label{371a}\\
\omega_r = U_0 k . \label{371b}
}{371}
The meaning of \equas{371} is the following. Time growing disturbances $(\omega_i > 0)$ are possible only if $R > k^2$. If $R < k^2$, all disturbances are damped in their time evolution $(\omega_i < 0)$. The marginal stability condition is simply
\eqn{
R = k^2 .
}{372}
Thus, the critical values of $k$ and $R$ for the onset of convective instability are obtained from minimising $R$ with respect to $k$ in \equa{372},
\eqn{
k_c = 0 , \quad R_c = 0 .
}{373}
This is the information that one can gather from \equa{371a}. A sketch of the transition from stability to convective instability in the plane $(k, R)$ is given in Fig.~\ref{fig4}. On the other hand, \equa{371b} makes it evident that a localised wave packet of temporal modes has a group velocity
\eqn{
\adervo{\omega_r}{k} = U_0 .
}{374}
and no spreading, since $d^2 \omega/ d k_r^2 = 0$. Then, the exponential time growth of the perturbation is possible only along the ray
\eqn{
\frac{x}{t} = U_0 .
}{374}
As a consequence, restriction of the analysis to temporal modes reveals the transition to convective instability, but does not allow to detect any possible condition of absolute instability, unless $U_0 = 0$. In this special case, the basic solution evolves directly from stability to absolute instability.

\subsection{Spatial modes}
\begin{figure}[t]
\begin{center}
\includegraphics[width=0.85\textwidth]{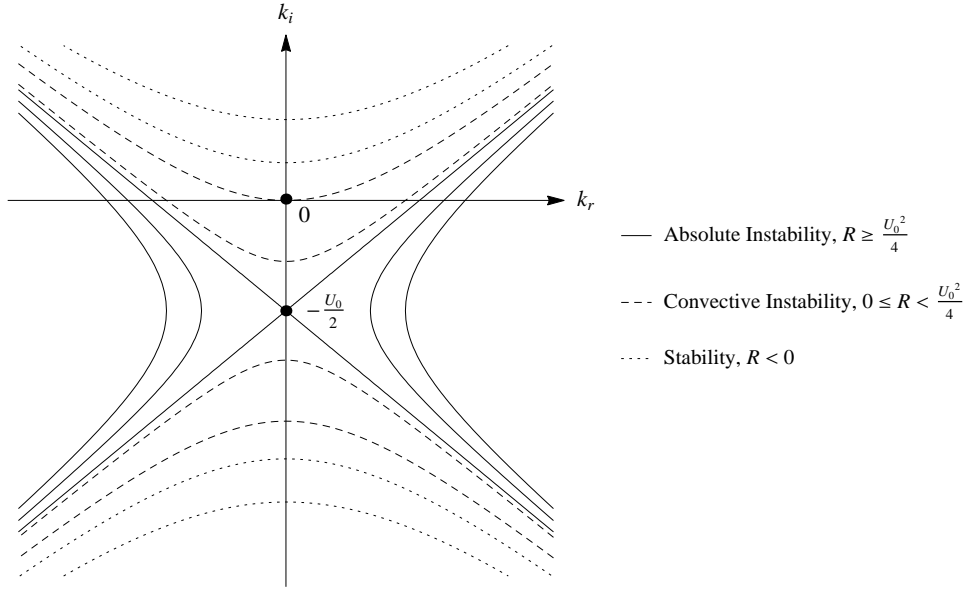}
\end{center}
\caption{Spatial modes: transition from stability to convective and absolute instability (with $U_0 > 0$)}
\label{fig5}
\end{figure}

In order to check if absolute instability is possible with $U_0 \ne 0$, we carry out a spatial stability analysis, then $k = k_r + i k_i \in \mathbb{C}$ and $\omega \in \mathbb{R}$. Separation of the real part from the imaginary part in \equa{37} yields
\gat{
k_r^2 - k_i^2 - U_0 k_i - R = 0, \label{38a}\\
\omega - U_0 k_r - 2 k_r k_i = 0 . \label{38b}
}{38}
\equa{38a} can be rewritten as
\eqn{
k_r^2 - \left(k_i + \frac{U_0}{2} \right)^2 = R - \frac{U_0^2}{4} .
}{39}
In this form, it is evident that \equa{39} describes a family of hyperbolae in the $(k_r, k_i)$--plane with asymptotes given by
\eqn{
k_r - k_i - \frac{U_0}{2} = 0, \qquad k_r + k_i + \frac{U_0}{2} = 0 .
}{40}
A sketch of these hyperbolae for a fixed $U_0 > 0$ and varying $R$ is provided in Fig.~\ref{fig5}.
The axis of each hyperbola is horizontal when $R < U_0^2/4$, namely $k_i = -\, U_0/2$, and vertical when $R > U_0^2/4$, namely $k_r = 0$. If $R < 0$, one branch of the hyperbola lies entirely in the upper half of the $(k_r, k_i)$--plane, i.e. $k_i > 0$, while the other branch lies in the lower half of the $(k_r, k_i)$--plane, i.e. $k_i < 0$. If $R=0$, the upper branch has a minimum coincident with the origin $(k_r = 0, k_i = 0)$, and thus is tangent to the $k_r$--axis. This threshold, $R=0$, defines the critical value of $R$, \equa{373}, and the onset of convective instability. For $R < R_c = 0$ we have two disconnected branches of modes, one growing downstream but damped upstream $(k_i < 0)$, and one growing upstream but damped downstream  $(k_i > 0)$. 

When the convective instability initiates, i.e. for $R > R_c = 0$, the upper branch crosses the $k_r$--axis thus including modes which grow downstream, while the lower branch raises upwards as $R$ increases. When $R = U_0^2/4$, the two branches merge at the pinching point $(k_r = 0, k_i = -\, U_0/2)$ and the hyperbola degenerates into a pair of straight lines given by \equa{40}. Eventually, when $R > U_0^2/4$, each branch of the hyperbola which now has a vertical axis $(k_r = 0)$ includes all the possible positive or negative spatial growth rates $k_i$, thus determining absolute instability.
 
We can write explicitly the threshold condition of absolute instability as
\eqn{
k = k_a = -\, i\, \frac{U_0}{2}, \qquad \omega = \omega_a = 0 , \qquad R = R_a = \frac{U_0^2}{4} ,
}{41}
where the value of $\omega_a$ is determined from \equa{38b}.

The group velocity $v_g = d \omega / d k$ for spatial modes with given $R$ and $U_0$ can be determined directly from \equa{37},
\eqn{
\aderv{}{k}\ \mathcal{D}(k, \omega ) = 0 \quad \Longrightarrow \quad i \left( \adervo{\omega}{k} - U_0 \right) - 2 k = 0 .
}{42}
We immediately infer from \equa{42} that $v_g = 0$ yields 
\eqn{
k = k_a = -\, i\, \frac{U_0}{2} .
}{43}
Thus, the condition $v_g = 0$ yields the pinching point in the $(k_r, k_i)$--plane defined by \equa{41}.

\section{Why spatial modes are necessary?}
The analysis of the sample case proposed in Section~\ref{burg} might give the impression that the temporal stability analysis and the spatial stability analysis give discrepant results relative to the transition to absolute instability. This impression could be the source of doubts about the correct path for a reliable analysis of this matter.

\subsection{General perturbations}
The first point is that the temporal modes are suitable to model a wave packet expressed as
\eqn{
F(x,t) = \int_{-\infty}^{\infty} \tilde{F}(k,t)\ e^{i k x} \ dk .
}{44}
We note that \equa{44} can be written whenever $F(x,t)$ satisfies the conditions for the existence of  the Fourier transform, namely of the integral
\eqn{
\tilde{F}(k,t) = \frac{1}{2 \pi} \int_{-\infty}^{\infty} F(x, t)\ e^{-i k x} \ dx .
}{45}
In general, the Fourier integral defined by \equa{45} does not exist, even according to the theory of distributions or generalised functions, if the given $F(x, t)$ is such that 
\eqn{
\lim_{x \to \pm\, \infty} |F(x,t)| = \infty .
}{46}
Thus, the point is: do we really need to envisage perturbations of the base flow state described by wave packets which satisfy \equa{46}? The answer is positive, as the perturbation wave packet does not have necessarily its source in a position close to the measurement station. The perturbation may be originated very far upstream or downstream, so that the distance to be covered by the wave packet along the $x$--direction to reach the measurement probe can be very large. Thus, an unstable perturbation wave packet might reach the probe when the amplitude at its source position is so large as to be considered infinite. This is the physical meaning of the condition expressed by \equa{46}, and this is the reason why the set of temporal normal modes is too tiny to include all possible disturbances acting on a base flow. 

\subsection{From the Fourier integral to the Laplace integral}

Now, we have a mathematical problem. We must replace \equa{44} with an integral expression compatible with \equa{46}. Unfortunately, we cannot achieve this task if we do not set some restrictions. Actually, we substitute \equa{46} with a much weaker condition, namely
\eqn{
\lim_{x \to \pm\, \infty} e^{- c |x|} |F(x,t)| = 0 , \qquad \mbox{for} \quad c \in \mathbb{R}, \ c > 0 .
}{47}
Therefore, \equa{44} can be replaced by
\eqn{
F(x,t) = 
\begin{cases}
\displaystyle\int_{-\infty - i \delta}^{\infty - i \delta} \tilde{F}^{(-)}(k,t)\ e^{-i k x} \ dk , \quad x<0,
\\[15pt]
\displaystyle\int_{-\infty - i \gamma}^{\infty - i \gamma} \tilde{F}^{(+)}(k,t)\ e^{i k x} \ dk , \quad x \geqslant 0 .
\end{cases}
}{48}
In \equa{48}, we denoted with $\delta$ and $\gamma$ the lower bounds of all the possible values of $c$ allowing \equa{47} to be satisfied for either $x<0$ or $x \geqslant 0$. Equation (\ref{48}) expresses $F(x,t)$ as a Laplace transform, where the Laplace variable is either $s=-ik$, when $x<0$, or $s=ik$, when $x \geqslant 0$.

\section{Conclusions}
This short report was aimed to propose a conceptual description of the transition from convective to absolute instability, as well as to give a reasonable and simple explanation of the need to go beyond the usual temporal stability analysis. The use of spatially growing modes along the direction of the base flow allows one to detect features of the instability that possibly remain hidden in the usual temporal stability approach. The latter is in fact perfectly effective when dealing with the transition from linear stability to convective instability, i.e. with the determination of the marginal stability condition and of the critical values of the governing parameters. On the other hand, temporal modes may prove to be ineffective in detecting the transition from convective to absolute instability. This possibility may be ascribed to those cases where the unstable behaviour, as detected in the laboratory reference frame, is activated preferably by perturbations whose source is localised far away from the measurement station. In order to establish these basic facts, the linear stability analysis of a toy flow system, governed by the one-dimensional Burgers' equation, was employed to provide an illustrative case study.

\bibliographystyle{ieeetr}
\bibliography{sample}

\end{document}